\documentclass[aps,prl,twocolumn,nofootinbib,superscriptaddress]{revtex4}%
\usepackage{color}
\usepackage{epsfig}
\usepackage{graphicx}
\usepackage{amsmath}
\usepackage{amsfonts}
\usepackage{amssymb}
\usepackage{epstopdf}
\usepackage{bbold}%
\providecommand{\U}[1]{\protect\rule{.1in}{.1in}}
\def\be{\begin{equation}}
\def\ee{\end{equation}}

\begin{document}
\title{Relative Entropy and Proximity of Quantum Field Theories}
\author{Vijay Balasubramanian}
\affiliation{David Rittenhouse Laboratories, University of Pennsylvania, Philadelphia, USA}
\affiliation{CUNY Graduate Center, Initiative for the Theoretical Sciences, New York, USA}
\affiliation{Theoretische Natuurkunde, Vrije Universiteit Brussel, and International Solvay Institutes, Pleinlaan 2, B-1050 Brussels, Belgium}
\author{Jonathan J. Heckman}
\affiliation{Department of Physics, University of North Carolina at Chapel Hill, USA}
\author{Alexander Maloney}
\affiliation{Department of Physics, McGill University, Montreal, Canada}

\begin{abstract}
We study the question of how reliably one can distinguish two
quantum field theories (QFTs). Each QFT defines a probability distribution on the space of fields.  The relative entropy provides a notion of proximity between these distributions and quantifies the number of measurements required to distinguish between them.
In the case of nearby conformal field theories,
this reduces to the Zamolodchikov metric on the space of couplings.
Our formulation quantifies the information lost under renormalization group flow
from the UV to the IR and leads us to a quantification of fine-tuning. This formalism
also leads us to a criterion for distinguishability of low energy effective field theories
generated by the string theory landscape.
\end{abstract}
\maketitle

\vspace{-3mm}

\section{Introduction}

\vspace{-3mm}

A central aim of high energy physics is to determine the operator content,
correlation functions, and coupling constants of the real world. This
problem is challenging,  especially in the context of string theory, because
there are a priori many UV completions of a given low energy effective field theory such as the
Standard Models of particle physics and cosmology.

Here we ask: given two competing theories of the world $p$ and $q$,
how reliably can we distinguish them given a finite number of  measurements?

Broadly speaking, theory determination is a basic question in statistical inference and information theory. By interpreting the (Euclidean continuation) of  a quantum field theory as a probability distribution on the space of field configurations, we shall convert well-studied information theoretic notions of proximity between probability distributions into analogous measures of proximity between QFTs.

This formalism gives a concrete method for evaluating the proximity
of QFTs in any UV complete theory with a landscape of low energy effective field theories.
Additionally, it provides a way to coarse-grain for specific features, and to quantify
the distinguishability of different effective field theories.

\vspace{-3mm}

\section{Proximity in Quantum Field Theory}

\vspace{-3mm}

One way to quantify the proximity\ between two probability distributions
$p(z)$ and $q(z)$ on a probability space is via the relative entropy (also called the
Kullback-Leibler (KL)\ divergence) \cite{KLDiv}:%
\begin{equation}
D_{KL}(p||q)=\int d\mu(z)\text{ \ }p(z)\log\frac{p(z)}{q(z)}.
\end{equation}
Here, $z$ is the outcome, and $d\mu(z)$ is a choice of measure on the space of
outcomes. The relative entropy $D_{KL}(p||q)\geq0$ and vanishes if and only if
$p=q$ almost surely.

In information theory, the KL\ divergence quantifies the amount of information
which is lost when one uses the distribution $q(z)$ to model the distribution $p(z)$.
In the context of statistical inference, one can consider $N$ independent
events $E=\{e_{1},...,e_{N}\}$ drawn from the distribution $q(z)$. At
large $N$, the probability that these draws could have been obtained from $p(z)$ is:
\begin{equation}
\text{Pr}(p|E)\simeq \exp(-ND_{KL}(p||q)).
\label{prob}
\end{equation}
For additional discussion and references to the literature, see e.g.
\cite{Balasubramanian:1996bn, Amari, CoverThomas}.

As it is not symmetric, the KL divergence is not really a distance.
Nevertheless, in the limit where $p$ and $q$ are nearby, the
KL divergence reduces to a metric.\footnote{This local metric also
arises from a more general notion of proximity obtained by taking the expectation value with
respect to the distribution $q$ of $f(p/q)$ for $f$ a convex function such
that $f(1)=0$ \cite{Chentsov}. The special case $f(u)=u\log u$ corresponds to the
KL\ divergence.
}
Consider a parametric family of distributions $q(z|\{\xi\})$ such that for
some value of $\xi$, $q(z|\{\xi_{\ast}\})=p(z)$. Expanding $D_{KL}(p
\vert\vert q)$ around this point yields the Fisher information metric:
\begin{align}
D_{KL}(p||q)  &  \simeq G_{MN}^{\mathrm{Fisher}}\text{ }\delta\xi^{M}\delta
\xi^{N}\\
G_{MN}^{\mathrm{Fisher}}  &  \equiv\int d\mu(z)\text{ \ }q(z)\frac
{\partial\log q}{\partial\xi^{M}}\frac{\partial\log q}{\partial\xi^{N}}.
\end{align}

In this note we will consider
probability distributions generated by a
Euclidean quantum field theory with action $S[\phi]$ depending on
field configurations $\{ \phi(x) \}$.   The action of the Euclidean field theory defines an
(unnormalized) probability distribution $\exp(-S[\phi])$.
This distribution defines the quantum field theory via  its analytic continuation to Lorentzian signature.
The normalized probability distribution on the space of field configurations $\{\phi(x)\}$ is %
\begin{equation}
p[\phi]=\frac{1}{\mathcal{Z}_{p}}\exp(-S_{p}[\phi])\text{ \ \ with
\ \ }\mathcal{Z}_{p}=\int\mathcal{D}\phi\text{ }\exp(-S_{p}[\phi]).
\end{equation}
Here $\mathcal{Z}_{p}$ is the partition function.
A draw from this probability distribution is a Euclidean field configuration $\phi(x)$.

Note that in quantum physics we are often also interested in a different distribution -- i.e. the square of the ground state wavefunction over equal time configurations in the Lorentzian theory.  The ground state wavefunction for a given spatial configuration $\Psi(\phi_0)$  is generated by summing the distribution $\exp(-S[\phi])$ over Euclidean trajectories that approach the boundary condition $\phi_0$ on a fixed Euclidean time surface.   We will {\it not} here be studying the associated distribution $|\Psi(\phi_0)|^2$, which is  also a quantity of  physical interest.  Our object of interest -- the probability distribution for $\phi(x)$ in Euclidean signature -- is most directly interpreted in statistical field theory rather than Lorentzian quantum field theory.

Given two quantum
field theories which depend on the same class of field configurations, we can
now study the KL\ proximity between two theories with actions
$S_{p}[\phi]$ and $S_{q}[\phi]$:%
\begin{equation}
D_{KL}(p||q)=\int\mathcal{D}\phi\,{\frac{e^{-S_{p}}}{\mathcal{Z}_{p}}}\left(
(S_{q}-S_{p})+\log\frac{\mathcal{Z}_{q}}{\mathcal{Z}_{p}}\right)  ,
\label{klqft}
\end{equation}
This is the expectation value of $(S_{q}-S_{p}%
)+\log\mathcal{Z}_{q}/\mathcal{Z}_{p}$ in the ground state of theory $p$.\footnote{For a wholly different discussion on the
distance between theories, see e.g. \cite{Calmet}.}

Note that this $D_{KL}$ is not the same thing as the quantum relative entropy,  Tr$(\rho_{p}\log\rho_{p}-\rho_{p}\log\rho_{q})$, between the ground states of theories $p$ and $q$,  where $\rho_i$ is
the density matrix for the ground state of theory $i$.  Nor is $D_{KL}$
the quantum relative entropy between two density matrices of a given theory.

\vspace{-3mm}

\subsection{Master Theories}

\vspace{-3mm}

At first glance, our notion of proximity would appear to only work for comparing QFTs with the same field / operator content. All that is
really required, however, is that there is some  \textquotedblleft master
UV\ theory\textquotedblright\ $p_{\text{master}}[\phi_{\text{master}}]$. This master theory
could be either a lattice formulation of a field theory, a continuum CFT, or a particular string compactification.

From this master theory we can consider deformations to various low energy effective field theories $q_{1},...,q_{M}$. Since all of the $q$'s descend from the same master theory, we can continue to label the field and operator content according to that of theory $p_{\text{master}}$.
Hence, we can still speak of the KL divergence $D_{KL}(q_{i}|| q_{j})$ for all $i$ and $j$. This point
will be especially important when we turn to the study of effective field theories generated by the string landscape.

\vspace{-3mm}

\subsection{Perturbative Calculability}

\vspace{-3mm}

The notion of proximity we have introduced is calculable in perturbation
theory. Consider a Euclidean theory with action $S_{p}[\phi]$, which we
perturb by a linear combination of local sources for operators:%
\begin{equation}
\Theta_{\lambda}(x)\equiv\lambda^{i}(x)O_{i}(x).
\end{equation}
Each $\lambda^{i}(x)$ specifies a source, i.e. a position dependent coupling
constant, although we will primarily focus on the case where the $\lambda^i$ are constant.
We then get a family of probability distributions $q[\phi|\left\{
\lambda\right\}  ]$ defined by the deformation:%
\begin{equation}
S_{q}[\phi] - S_{p}[\phi] =\int d^{D}x\sqrt{g_{(x)}}\text{ }\Theta_{\lambda}(x).
\end{equation}
This deformation is proportional to the trace of the difference in stress energies between
theories $p$ and $q$:
\begin{equation}
T_{\mu\nu}^{(q)}-T_{\mu\nu}^{(p)}=\frac{2}{\sqrt{g_{(x)}}}\frac{\delta
(\sqrt{g_{(x)}}\text{ }\Theta_{\lambda}(x))}{\delta g_{(x)}^{\mu\nu}}%
=g_{\mu\nu}\Theta_{\lambda}(x)\text{.}%
\end{equation}
The KL proximity between $p$ and $q$ is calculable in perturbation theory since the term  $S_q - S_p$ depends only on the expected value of the deformation, while the partition functions ${\cal Z}_p,{\cal Z}_q,$ are the generating functions of the correlation functions of $p$ and $q$.  Operationally, we do not even need an action for either theory, but only
their correlation functions. In this sense the KL divergence
also quantifies the amount of information contained
in the correlation functions of a theory.

We now study the leading order behavior of the KL\ divergence,
with $\Theta_{\lambda}(x)$ treated as a small perturbation to the original
theory. Expanding $q[\phi]$ to quadratic order in the perturbation, the
KL\ divergence is
\begin{equation}
D_{KL}(p||q)=\frac{1}{2}\int d^{D}x\sqrt{g_{(x)}}d^{D}y\sqrt{g_{(y)}%
}G_{\lambda}(x,y)^{\text{conn}}+...,
\label{KLZ}
\end{equation}%
where $G_{\lambda}(x,y)^{\text{conn}}$ is the connected two point function
\begin{equation}
G_{\lambda}(x,y)^{\text{conn}}\equiv\left\langle \Theta_{\lambda}%
(x)\Theta_{\lambda}(y)\right\rangle _{p}-\left\langle \Theta_{\lambda
}(x)\right\rangle _{p}\left\langle \Theta_{\lambda}(y)\right\rangle _{p}.
\end{equation}
We conclude that the Fisher information metric on the space of couplings is
\begin{equation}
G_{ij}^{Fisher} = \frac{1}{2}\int d^{D}x\sqrt{g_{(x)}}d^{D}y\sqrt{g_{(y)}}
\left(
\left\langle O_i O_j\right\rangle_{p} - \langle O_i \rangle_{p} \langle O_j\rangle_{p}
\right).
\end{equation}

Integrated two-point functions of this sort have two types of divergences -- an IR divergence proportional to the volume and a UV divergence coming from contact terms where $x$ and $y$ coincide.  The IR divergence is easily regulated by putting the system at finite volume.  The UV divergent contact terms are generally scheme dependent. If a particular finite UV completion is known, this fixes the regularization scheme unambiguously.
As one might expect, the largest contribution to the KL divergence comes from such UV divergent pieces.

The finite part of $D_{KL}(p||q)$ is independent of the UV cutoff $\Lambda_{UV}$ and can be
packaged in terms of the values of the couplings at an RG scale $\mu$.
Operationally, we introduce a regularization scheme, along
with some counterterms in theory $p$. This finite piece is independent of $\Lambda_{UV}$, but can
depend on a choice of scheme, and can a priori be positive, zero, or negative.  This makes the physical interpretation of the finite piece of  $D_{KL}(p||q)$ more subtle.
As we change $\Lambda_{UV}$,
the values of the couplings at $\mu$ must be adjusted to hold fixed the long distance behavior.
This active tuning of the couplings is reflected in the fact that the beta function:
\begin{equation}
\beta_{KL} \equiv \frac{\partial D_{KL}(p||q)}{\partial \log \mu}.
\end{equation}
is independent of $\Lambda_{UV}$ and can a priori be positive, zero, or negative.

An important special case is when $p$ and $q$ are related by renormalization group flow; they describe a given quantum field theory with UV momentum cutoffs $\Lambda_{UV}^{(p)}$ and $\Lambda_{UV}^{(q)}$, respectively.  We will take $\Lambda_{UV}^{(q)} < \Lambda_{UV}^{(p)}$.
We can regard theories $p$ and $q$ as defining different probability measures on the same configuration space, in the usual way:
one first integrates out field configurations $\phi_k$ in theory
$p$ with momenta $\Lambda_{UV}^{(q)}<|k|<\Lambda^{(p)}_{UV} $ and then rescales positions and momenta as $x\rightarrow b x$ and $k\rightarrow
k/b$, where $b=\Lambda_{UV}^{(q)}/\Lambda^{(p)}_{UV}$.
This defines two distributions with different coupling constants related by renormalization group flow.
The KL divergence then provides a measure of the information lost as one
coarse grains from $\Lambda_{UV}^{(p)}$ to $\Lambda_{UV}^{(q)}$.\footnote{Another quantity of interest is the mutual information. Starting with a master theory $p_{\text{master}}[\Lambda_{IR} , \Lambda_{UV}]$ with IR and UV cutoffs $\Lambda_{IR}$ and $\Lambda_{UV}$, marginalize out either high or low momentum shells, to respectively produce distributions $p_{hi}[\Lambda_{IR} , \mu]$, and $p_{lo}[\mu , \Lambda_{UV}]$. The second operation is somewhat awkward in local quantum field theory, but makes sense both in the context of non-commutative field theories, and in theories which have a gravity dual with a finite length AdS throat. The product distribution $p_{hi} \times p_{lo}$ has support on the same momentum modes as $p_{\text{master}}$, and $D_{KL}(p_{\text{master}} || p_{hi} \times p_{lo}) = \mathcal{I}(UV , IR)$ is the mutual information between the UV and IR. For related discussions of the relations between renormalization group flow and information theory see \cite{Gaite:1995yg, Brody:1997gn, Casini:2006es, Apenko:2009kq, Balasubramanian:2011wt}, as well as \cite{Beny:2012, Beny:2013rea}. For further discussion on information geometry in the context of AdS / CFT, see e.g. \cite{Blau:2001gj, Rey:2005cn}.}

\vspace{-3mm}

\section{Conformal Field Theories}

\vspace{-3mm}

We now consider the special limit where either theory $p$ or theory $q$ is
a conformal field theory (CFT).

First, suppose that $p$ is a CFT and that $q$ is
another CFT obtained by perturbing $p$ by a linear combination of exactly marginal scalar operators.
The small variation $\delta\lambda^{i}$ can be viewed as a vector in the
space of marginal couplings.
In our conventions, the two-point function for a
marginal primary scalar of dimension $\Delta=D$ is %
\begin{equation}
\left\langle O_{i}(x)O_{j}(0)\right\rangle _{p}=G_{ij}^{(\text{Zam})}\frac
{1}{(2\pi)^{D}}\frac{1}{\left\vert x\right\vert ^{2D}},
\end{equation}
where $G_{ij}^{(\text{Zam})}$ is the Zamolodchikov metric of the CFT
\cite{Zam}.  Then, since the CFT one-point functions vanish, the KL divergence (\ref{KLZ})
is proportional to the length of the vector $\delta \lambda^i$ with respect to the Zamolodchikov metric:
\begin{equation}
\delta\lambda = \sqrt{\delta \lambda^i \delta \lambda^j G_{ij}^{(\text{Zam})}}
\end{equation}
Thus the Fisher information metric is proportional to the Zamolodchikov metric!

To make this more explicit, consider theories defined on a lattice of volume $V = \ell_{IR}^D$ with lattice-separation $\ell_{UV}$.  Then
\begin{equation}
D_{KL}(p||q)\simeq \delta \lambda^2 \times\left(  \frac{V}{\ell_{UV}^D}\right).
\end{equation}
The distance $\delta \lambda$ in the space of couplings can then be interpreted as the KL density, as follows.
Since there are $K = V/\ell_{UV}^D$ lattice sites, each draw from the Euclidean probability distribution gives $K$  pieces of data about the couplings of the theory. Following (\ref{prob}),  a measurement of the field configuration at one lattice site will fail to distinguish between the two theories with probability $e^{-\delta \lambda^2}$.

From the perspective of a continuum theory, the lattice  described above is a
particular regularization scheme, which will contain UV divergences as the lattice cutoff
is taken to zero size. It is also of interest to extract the
finite piece which remains in the continuum limit.
In a CFT, different choices of regularization scheme correspond to different choices of contact terms in the OPE.  Naively, these might appear to change the Zamolodchikov metric.  However, changes of scheme can be interpreted as coordinate transformations on the space of couplings $\lambda^i$; the Zamalodchikov metric transforms covariantly under these diffeomorphisms (see e.g. \cite{Kutasov:1988xb}).
The scheme-independent piece of $D_{KL}$ is, up to a factor of order one, proportional to
the Zamolodchikov distance in the space of couplings:
\begin{equation}
D_{KL}(p||q) \propto   \delta \lambda^2.
\label{KLIR}
\end{equation}
The factor of $V/\ell_{UV}^D$ has disappeared, since the regulated integrals appearing in (\ref{KLZ}) must be proportional to $\ell^D_{UV}/V$ for dimensional reasons. We can interpret this as follows: the finite piece of $D_{KL}$ is not proportional to the volume $V$ because there are long range correlations in a conformal field theory, and hence a given draw from the distribution essentially gives one piece of data about the theory.  The precise coefficient in (\ref{KLIR}) depends on the nature of the IR regularization.\footnote{For some examples of
such computations for CFTs on round spheres in various dimensions, see e.g. \cite{Gerchkovitz:2014gta}.}

Let us now consider the case when the theory $q$ is not a CFT, but is related to theory $p$ by the addition of some non-marginal operators $O_i$.  We will work in a basis where $G_{ij}^{(\text{Zam})}$ is diagonal.  Then when the perturbations $\delta \lambda^i$ are small the above derivation can be easily generalized. In the lattice regularization
\begin{equation}\label{CDEF}
D_{KL}(p||q)\simeq\sum_i c_i~ \left(\delta\lambda^i\right)^2  \ell_{IR}^{2(D-\Delta_i ) } \left(\frac{\ell_{IR}}{\ell_{UV}}\right)^{2 \Delta_i-D}
\end{equation}
where the $c_i$ are numerical constants of order one.

For each summand in equation (\ref{CDEF}), the finite piece of $D_{KL}(p||q)$ is proportional to $c_i~ \left(\delta\lambda^i\right)^2  \ell_{IR}^{2(D-\Delta_i ) }$. As expected, the contribution to $D_{KL}(p||q)$ is dominated by the contribution of the lowest dimension (most relevant) operator when $\ell_{IR}$ is large.   Irrelevant couplings ($\Delta > D$) make a finite contribution to $D_{KL}$ that is suppressed by the infrared scale because the low-energy effective theories are identical.   Nearly marginal perturbations ($\Delta \sim D$) contribute to $D_{KL}$ in a way that is almost insensitive to volume because nearly conformal theories have long-range spatial correlations and hence measurements at different locations do not give independent information about the theory.    Relevant perturbations with dimensions above the Breitenlohner-Freedman bound ($D > \Delta > D/2$) all lead to sub-extensive scaling of $D_{KL}$ with volume, but the unitarity bound $\Delta > D/2 - 1$ leaves a narrow window with super-extensive scaling. It would be interesting to understand how this arises in terms of measurements distinguishing $p$ from $q$.

Conversely, suppose $q$ is an IR fixed point, and $p$ is nearby. Similar statements apply,
since to leading order $D_{KL}(p||q)=D_{KL}(q||p)+O(\delta\lambda^{3})$.

\vspace{-3mm}

\subsection{Renormalization Group Flows}

\vspace{-3mm}

Let us now consider the case where the deformation initiates an
RG flow from the UV theory $p$ to the IR theory $q$. The resulting
flow and subsequent form of the KL divergence will be
dominated by the operator(s) of lowest dimension $\Delta<D$.
In the special case where the operator is marginally relevant, i.e. has dimension
$\Delta=D-\delta$ for $\delta\ll1$, this flow is short. In many situations
such as 2D\ minimal models with central charge close to one,
and various 4D\ supersymmetric quantum field theories, $\delta$
is calculable. The KL\ divergence in this case can again be computed and we
get (in a lattice regularization) precisely (\ref{KLIR}) described above.

We can also consider the KL divergence between two points along an RG flow,
$D_{KL}(t_p|| t_q)$ as a function of the RG flow parameter $t=\log \mu$.
Specializing to the case of a 2D\ CFT, we learn that the initial change in the
central charge is closely related to the information lost in moving from the
UV to the IR.  We have
\begin{equation}
\left.{\partial^2 D_{KL}(t||t_q)\over \partial t^2}\right|_{t_q=t} \propto G_{ij}^{(\text{Zam})}{\partial \lambda^{i}\over \partial t}{\partial\lambda^{j}\over \partial t}%
=-12\frac{\partial c(t)}{\partial t},
\end{equation}
where $c(t_{RG})$ is the c-function of a two-dimensional conformal field
theory along the flow \cite{Zam}.

\vspace{-3mm}

\subsection{Metric Proximity}

\vspace{-3mm}

More generally, there is a deep intuition that the conformal anomalies of a CFT measure its
degrees of freedom. We now show that this statement has a sharp information-theoretic interpretation.
Consider a Euclidean signature conformal field theory on a $D$%
-dimensional manifold $M_{D}$. Varying the background metric $g_{\mu\nu}$
defines a family of theories $p[\phi|\{g_{\mu\nu}\}]$, and it is natural to
consider the proximity of two such members. Perturbing about a fixed
background $g\rightarrow g+\delta g$, the line element for the information
metric is:
\begin{equation}
\frac{1}{2} \int d^{D}x\sqrt{g_{(x)}}d^{D}y\sqrt{g_{(y)}}\left\langle \mathcal{T}^{\mu\nu
}(x)\mathcal{T}^{\rho\sigma}(y)\right\rangle _{p}\delta g_{\mu\nu}(x)\delta
g_{\rho\sigma}(y)
\end{equation}
where $\mathcal{T}^{\mu\nu}(x)=T^{\mu\nu}(x)-\left\langle T^{\mu\nu
}(x)\right\rangle _{p}$ is the stress energy tensor with the one-point
function subtracted off. For $D$ odd this one-point function vanishes, and for
$D$ even, it is determined by the conformal anomaly.

Evaluating on $M_{D}$ conformally equivalent to flat space, the two-point
function for $\mathcal{T}^{\mu\nu}$ is closely related to the evaluation of a
particular linear combination of central charges which counts the local degrees of freedom in the field
theory. Recall that in flat space, we have:
\begin{equation}
\left\langle \mathcal{T}^{\mu\nu}(x)\mathcal{T}^{\rho\sigma}(0)\right\rangle
_{\mathbb{R}^{D}}=\left\langle T^{\mu\nu}(x)T^{\rho\sigma}(0)\right\rangle
_{\mathbb{R}^{D}}\equiv C_{T}\frac{I^{\mu\nu,\rho\sigma}(x)}{x^{2D}},
\end{equation}
where $I^{\mu\nu,\rho\sigma}(x)$ is a specific dimensionless combination of
terms quadratic in the positions, as dictated by conformal invariance (see
e.g. \cite{Osborn:1993cr}). Here, $C_{T}>0$ in a reflection positive theory,
which agrees with the information theoretic condition $D_{KL}(p \vert\vert q)
\geq0$. In two and four dimensions, $C_{T}$ is proportional to $c$, and in
three-dimensional $\mathcal{N}=2$ supersymmetric field theories it is
proportional to $\tau_{RR}$, i.e. the normalization constant for the
R-symmetry current two-point function (see e.g. \cite{Barnes:2005bm}).

This is a satisfying result. It tells us that the quantity $C_T$ is proportional to $D_{KL}$,
directly quantifying the level of distinguishability encoded in local degrees of freedom.

We can also extend this calculation to cover the case of RG flows. Along these lines, we introduce a UV cutoff $\Lambda_{UV}$, and
consider two UV CFTs which differ only in the choice of background metric $g_{\mu \nu}$ and a small perturbation to
another metric $g_{\mu \nu} + \delta g_{\mu \nu}$.
Suppose we perturb this UV CFT by a relevant operator. Upon flowing to the IR,\footnote{In a CFT with a UV cutoff, we can perform an RG flow by a Weyl rescaling of our background metric. We thank H. Verlinde for this comment.} we can evaluate
the KL proximity for these two background metrics. Hence, we see that if the two
theories are closer together in the IR, then $C_{T}^{UV} > C_{T}^{IR}$. So in
other words, the statement that $C_T$ typically decreases under RG flow
means CFTs typically get closer in the IR.

\vspace{-3mm}

\section{Landscapes}

\vspace{-3mm}

One clear lesson from recent work in string theory is the existence of a large landscape of self-consistent low energy
effective field theories. We now show how to deploy our formalism in the study of the landscape.

\vspace{-3mm}

\subsection{Flux Vacua}

\vspace{-3mm}

In flux vacua (see e.g. \cite{Douglas:2006es} for a review),
the flux quantum numbers define an integrally quantized
lattice vector $\overrightarrow{N}$, and with it an effective action
$S[ \varphi_1,...,\varphi_n , \overrightarrow{N}]$ for some fields $\varphi_1,...,\varphi_n$. Given two flux
vectors $\overrightarrow{N}$ and $\overrightarrow{M}$, we can compute the
proximity $D_{KL}(\overrightarrow{N} || \overrightarrow{M})$ between the
two effective theories. Note that a priori, this has nothing
to do with the distance between $\overrightarrow{N}$ and $\overrightarrow{M}$ on the
lattice of fluxes.

To illustrate, consider a toy model in which our
flux vector $\overrightarrow{N}$ generates
an effective action for a single canonically normalized real scalar $\phi$ with $l$ isolated vacua:
\begin{equation}
S[\phi , \overrightarrow{N}] = \int d^{4}x\text{ }\left(  \frac{1}{2}\left(  \partial\phi\right)
^{2} - \frac{1}{\Lambda^{2l-4}}\underset{k=1}{\overset{l}{%
{\displaystyle\prod}
}}(\phi-\phi^{(k)})^{2}
\right).
\end{equation} This form readily generalizes to complex scalars,
as well as supersymmetric models.

Suppose now that we have another flux vector $\overrightarrow{M}$ such that the form of the effective potential
for this flux vector has minima which are nearby the minima of the theory with flux vector $\overrightarrow{N}$. This means the KL
divergence can be evaluated by just varying the $l$ minima in the distribution
$p(\phi | \{\phi^{(1)} ,..., \phi^{(l)} \})$:
\begin{equation}
D_{KL}(\overrightarrow{N} || \overrightarrow{M}) \simeq G_{ij} \delta \phi^{(i)} \delta \phi^{(j)},
\end{equation}
where $G_{ij}$ is the information metric from varying with respect to the
locations of the minima. Working in a saddle-point approximation around
each of the $l$ massive vacua yields the approximation:
\begin{equation}
G_{ij} \simeq V \times \frac{\delta_{ij}}{2 l}  \times m^2_{i}
\end{equation}
where $m_{i}^2 = \frac{2}{\Lambda^{2l - 4}}\prod_{k \neq i} (\phi^{(k)} - \phi^{(i)})^2$ is
the mass-squared of the real scalar expanded around the $i^{th}$ critical point, and $V$ is the regulated volume of the
spacetime.

\vspace{-3mm}

\subsection{2D CFTs}

\vspace{-3mm}

2D CFTs are another machine for generating a vast number of vacua in (perturbative) string theory.
The Zamolodchikov metric is insufficient to define a notion of distance since it cannot connect all CFTs \cite{DougLand}. A formal notion of distance given in \cite{DougLand} centered on defining a metric on the values of local $n$-point functions. Though the details differ, evaluating the KL divergence intuitively agrees with this, since it involves integrated $n$-point functions.

Let us illustrate in more detail for 2D CFTs defined on a torus. Consider the two $c = 4/5$ theories, with
diagonal and off-diagonal partition functions, which respectively correspond to the tetra-critical Ising model and the three state Potts model.
Although they cannot be connected by an operator deformation, they both descend from the same UV spin system and so there ought to be a ``distance'' between these theories \cite{DougLand}. Following our general considerations, we take the UV spin system to define our
master theory $p_{\text{master}}$ with the two $c = 4/5$ models viewed as effective field theories $q_i$.

The value of the KL divergence strongly depends on the UV lattice spacing, and as we now argue, diverges as we pass to the continuum limit.
To see this, observe that the off-diagonal theory is a $\mathbb{Z}_2$ orbifold of the diagonal theory \cite{Schellekens:1989am}. Though the untwisted sector of the orbifold coincides with the singlet sector of the parent, the parent has non-singlet states, and the orbifold has twisted sector states. These additional states mean that some states of each theory are not present in the other, and so $D_{KL}$ is infinite in both directions.

Since we have a $\mathbb{Z}_2$ symmetry, it is also natural to compute the KL divergence between the singlet sector and the full theory. This is not really a direct comparison of the two $c = 4/5$ CFTs, but provides a way of telling us whether a theorist with restricted knowledge about their CFT could ascertain information about the full CFT. The KL divergence $D_{KL}(p_{full}||p_{sing})$ is still infinite, but in the other direction, we get:\footnote{The diagonal $c = 4/5$ model has primaries with
weights $h_{r,s} = ((6r - 5s)^2 - 1)/120$ for $r = 1,...,4$ and $s = 1,...,r$. In terms of the
Virasoro characters $\chi^{\pm}_{r,s}$,
$Z_{full} = |\chi^{+}_{1,1}|^2 + |\chi^{+}_{2,1}|^2 + |\chi^{-}_{2,2}|^2
+ |\chi^{+}_{3,1}|^2 + |\chi^{-}_{3,2}|^2 + |\chi^{+}_{3,3}|^2
+ |\chi^{+}_{4,1}|^2 + |\chi^{-}_{4,2}|^2 + |\chi^{+}_{4,3}|^2 + |\chi^{-}_{4,4}|^2 $.
We get $Z^{diag}_{sing}$ by omitting all
of the $\chi^{-}_{r,s}$ terms. The latter is a function on the Teichm\"{u}ller space
for the complex structure modulus of the $T^2$.}
\begin{equation}
D_{KL}(p^{diag}_{sing} || p^{diag}_{full}) = \log \left( \frac{Z^{diag}_{full}}{Z^{diag}_{sing}} \right).
\end{equation}
Similar considerations apply for the orbifold theory and its untwisted sector. Clearly,
our considerations extend to more general orbifold constructions.

\vspace{-3mm}

\subsection{Quantifying Fine-Tuning}

\vspace{-3mm}

To a low energy effective field theorist, what really matters is whether such UV completions lead to novel constraints
on IR physics. Starting from some UV master theory, we might imagine that upon an appropriate operator deformation,
there is a collection of intermediate values of the couplings, and corresponding theories
$p_{1},...,p_{M}$ which upon further flow respectively descend to
$q_{1},...,q_{M}$. Given a set of $M$ such RG\ trajectories, we can therefore
evaluate $D_{ij}^{(p)}=D_{KL}(p_{i}||p_{j})$ and $D_{ij}^{(q)}=D_{KL}%
(q_{i}||q_{j})$, and the corresponding ratios:
\begin{equation}
F_{ij}\equiv\frac{D_{ij}^{(p)}}{D_{ij}^{(q)}}. \label{FINT}%
\end{equation}
We say that a pair of theories is fine-tuned when $F_{ij}\ll1$. When
$F_{ij}\gg1$, then we say that the theory has no fine-tuning. Intermediate
cases can also be evaluated by a similar token.

\vspace{-3mm}

\subsubsection{$\phi^4$ Theory}

\vspace{-3mm}

As an illustrative example, consider the theory of a single real scalar with potential
$V(\phi) = m^2 \phi^2 / 2 + \lambda \phi^4 / 4!$. This theory is fine-tuned because
small perturbations in the UV boundary conditions of the coupling constants
lead to large changes in the IR parameters of the effective theory. Treating $m$ and $\lambda$ as bare parameters of a UV theory, we
can evaluate the $2 \times 2$ information metric in this case to find the leading order cutoff dependent contributions:
\begin{equation}
\left[
\begin{array}
[c]{cc}%
G_{m^{2}m^{2}} & G_{m^{2}\lambda}\\
G_{\lambda m^{2}} & G_{\lambda\lambda}%
\end{array}
\right]  \sim \left(\frac{\Lambda_{UV}}{\Lambda_{IR}} \right)^{4} \times\left[
\begin{array}
[c]{cc}%
1/\Lambda_{UV}^{4} & \lambda/\Lambda_{UV}^{2}\\
\lambda/\Lambda_{UV}^{2} & \lambda^{2}%
\end{array}
\right]  , \label{fishphri}%
\end{equation}
where each entry is multiplied by an \textquotedblleft order one
number\textquotedblright, as follows from dimensional analysis considerations. As expected, if we evaluate the proximity of two theories
in the IR, there is a power law divergent contribution. In the $\overline{MS}$ scheme,
the leading log contribution is:\footnote{For further details on the evaluation of
the three-loop basketball diagram entering in the
computation of $G_{\lambda \lambda}$, see \cite{Andersen:2000zn}.}
\begin{equation}
\left[
\begin{array}
[c]{cc}%
G_{m^{2}m^{2}} & G_{m^{2}\lambda}\\
G_{\lambda m^{2}} & G_{\lambda\lambda}%
\end{array}
\right]_{\overline{MS}}  =  V \times \left[
\begin{array}
[c]{cc}%
\mathcal{L} / 4 & - m^2 \mathcal{L}^2 / 8 \\
- m^2 \mathcal{L}^2 / 8 & 3 m^4 \mathcal{L}^3 / 16  %
\end{array}
\right]
\end{equation}
with $\mathcal{L} = \log(\mu^2 / m^2) / 16 \pi^2$, and $V$ the regulated volume of the spacetime.

\vspace{-3mm}

\subsection{Large Field Range Inflation}

\vspace{-3mm}

A common claim in the study of string compactifications is that since the inflaton is sensitive to Planck scale
physics, learning the exact shape of the inflaton potential would provide a wealth of information on the UV structure of a theory.
Here, we quantify the amount of information obtained from the first correction to the simple $m^2 \phi^2 / 2$ potential of large field
range inflation. We consider a correction term of order $\lambda \phi^4 / 4! $ and address the distinguishability of
the theory with $\lambda = 0$ versus $\lambda \neq 0$.

Along these lines, we return to our calculation for $\phi^4$ theory, viewing the reduced Planck scale $M_{PL} \sim 10^{18}$ GeV as a
UV cutoff, and $m$ as a soft IR cutoff. Using our methodology,
we get that the KL divergence scales as $G_{\lambda\lambda}$ in equation
(\ref{fishphri}):%
\begin{equation}
D_{KL}(\{\lambda=0\}||\{\lambda\neq0\})\sim\lambda^{2}\times\left(
\frac{M_{PL}}{m}\right)  ^{4}.
\end{equation}
On the other hand, to not spoil slow roll in the first place, we need to assume $\lambda < (m / \Delta \phi)^2$ so for a field
range $\Delta \phi \sim 10 M_{PL}$, we learn that the KL divergence is bounded above:
\begin{equation}
D_{KL}(\{\lambda=0\}||\{\lambda\neq0\}) < \left( \frac{M_{PL}}{\Delta \phi} \right)^4 \sim 10^{-4}.
\end{equation}
This upper bound is rather charitable, as it is the information
content over the entire volume of the spacetime.

\vspace{-3mm}

\section{Discussion}

\vspace{-3mm}

Viewing quantum field theory as a machine for generating probability
distributions on the space of fields, the relative entropy
leads to a measure of proximity in the space of
QFTs. In the special case of conformal field theories connected by marginal deformations,
we recover the familiar case of the Zamolodchikov metric. We have also seen how to track
information loss both in terms of RG flows and the value of $C_T$ (for CFTs).

Using this setup, we can coarse-grain any landscape of low energy
effective field theories. We simply ask how many
idealized measurements (i.e. draws of field configurations) must be performed before we can reliably
distinguish two theories. This dovetails with recent investigations aimed
at understanding how well a low energy observer could reconstruct --even in principle--
different UV completions \cite{StatString, HebInf}.

In future work, it would be interesting to study the behavior of the KL
divergence in various covariant regulator schemes such as \cite{CNC}, and also
to apply our formalism in various scenarios where operators of the Standard Model mix with an extra sector.
It would also be exciting to consider more general systems such as spin glasses, where
it is quite common to encounter statistical ensembles of coupling constants.


\textit{Acknowledgements:} JJH\ and AM\ thank T. Hartman for helpful
discussions and collaboration at an early stage of this work. We also thank R. Blumenhagen, T.
Dumitrescu, M. Freytsis, A. Hebecker, S. Hellerman, A. N. Schellekens, J. Sonner, and H. Verlinde
for helpful discussions and correspondence. JJH thanks
the high energy theory group at UPenn, and the organizers of the workshop ``Frontiers in String
Phenomenology'' for kind hospitality at Schloss Ringberg during part of this work. AM
thanks the Harvard high energy theory group for hospitality during his sabbatical leave.
JJH and AM thank the CUNY graduate center for hospitality during part of
this work. AM and VB thank the Aspen Center for Theoretical
Physics during part of this work. The work of VB is supported by DOE grant DE-FG02-05ER-41367,
and the work of AM is supported by an NSERC Discovery grant.

\vspace{-3mm}


\begin{thebibliography}{99}                                                                                               %

\bibitem {KLDiv}S.~Kullback, R.~A.~ Leibler,
Ann. of Math. Stat. \textbf{22}, No. 1, 79-86 (1951).

\bibitem{Balasubramanian:1996bn}
  V.~Balasubramanian,
  Neur. Comp. \textbf{9}, No. 2, 349-368 (1997), cond-mat/9601030.

\bibitem {Amari}S.-I.~Amari and H.~Nagaoka, ``Methods of Information
Geometry'', vol. 191 \textit{Translations of Mathematical Monographs; American
Mathematical Society and Oxford University Press}, Providence, RI (2000).

\bibitem {CoverThomas}T.~M.~Cover and J.~A.~Thomas, ``Elements of Information
Theory'', \textit{Wiley}, New York, NY (1991).

\bibitem {Chentsov}N.~N.~Chentsov, ``Statistical Decision Rules and Optimal
Inference'', \textit{Translations of Mathematical Monographs; American
Mathematical Society}, Providence, RI (1982).

\bibitem {Calmet}J.~Calmet and X.~Calmet, Mod. Phys. Lett. A {\bf 26}, 319 (2011).

\bibitem{Balasubramanian:2011wt}
  V.~Balasubramanian, M.~B.~McDermott and M.~Van Raamsdonk,
  Phys.\ Rev.\ D {\bf 86}, 045014 (2012), hep-th/1108.3568.


\bibitem{Casini:2006es}
  H.~Casini and M.~Huerta,
  J.\ Phys. {\bf A40}, 7031 (2007), cond-mat/0610375.

\bibitem{Brody:1997gn}
  D.~C.~Brody and A.~Ritz,
  Nucl.\ Phys. {\bf B522}, 588 (1998), hep-th/9709175.

\bibitem{Apenko:2009kq}
  S.~M.~Apenko,
  Physica {\bf A391}, 62 (2012), cond-mat/0910.2097.


\bibitem{Gaite:1995yg}
  J.~C.~Gaite and D.~O'Connor,
  Phys.\ Rev. {\bf D54}, 5163 (1996), hep-th/9511090.

\bibitem{Beny:2012}C.~Beny and T.~J.~Osborne,
  quant-ph/1206.7004.

\bibitem{Beny:2013rea}
  C.~Beny and T.~J.~Osborne,
  quant-ph/1310.3188.

\bibitem{Blau:2001gj}
  M.~Blau, K.~S.~Narain and G.~Thompson,
  hep-th/0108122.


\bibitem{Rey:2005cn}
  S.~J.~Rey and Y.~Hikida,
  JHEP {\bf 0608}, 051 (2006), hep-th/0507082.

\bibitem {Zam}A.~B.~Zamolodchikov,
JETP Lett \textbf{43}, No. 12, 730-732 (1986).

\bibitem{Kutasov:1988xb}
  D.~Kutasov,
  Phys.\ Lett.\ B {\bf 220}, 153 (1989).

\bibitem{Gerchkovitz:2014gta}
  E.~Gerchkovitz, J.~Gomis and Z.~Komargodski,
  hep-th/1405.7271.

\bibitem {Osborn:1993cr} H.~Osborn and A.~C.~Petkou,
Ann. Phys. \textbf{231}, 311 (1994), hep-th/9307010.


\bibitem {Barnes:2005bm}E.~Barnes, E.~Gorbatov, K.~A.~Intriligator, M.~Sudano
and J.~Wright,
Nucl.\ Phys.\ B \textbf{730}, 210 (2005), hep-th/0507137.

\bibitem{Douglas:2006es}
  M.~R.~Douglas and S.~Kachru,
  Rev.\ Mod.\ Phys.\  {\bf 79}, 733 (2007), hep-th/0610102.

\bibitem {DougLand}M.~R.~Douglas,
hep-th/1005.2779.

\bibitem{Schellekens:1989am}
  A.~N.~Schellekens and S.~Yankielowicz,
  Nucl.\ Phys.\ B {\bf 327}, 673 (1989).

\bibitem{Andersen:2000zn}
  J.~O.~Andersen, E.~Braaten and M.~Strickland,
  Phys.\ Rev.\ D {\bf 62}, 045004 (2000), hep-ph/0002048.



\bibitem {StatString}J.~J.~Heckman,
hep-th/1305.3621.

\bibitem {HebInf}A.~Hebecker,
Phys. Rev. \textbf{D88}, 125025 (2013), hep-th/1305.6311.

\bibitem{CNC}
  J.~Heckman and H.~Verlinde,
  Nucl.\ Phys. {\bf B894}, 58 (2015), hep-th/1401.1810.



\end{thebibliography}
\end{document}